\documentclass[aps,showpacs,prd,preprint]{revtex4}
\usepackage{graphicx}
\usepackage[all]{xy}
\usepackage{amsmath}
\usepackage{amssymb}
\newcommand{\be}{\begin{equation}}
\newcommand{\ee}{\end{equation}}
\newcommand{\ben}{\begin{eqnarray}}
\newcommand{\een}{\end{eqnarray}}
\newcommand{\bes}{\begin{subequations}}
\newcommand{\ees}{\end{subequations}}

\newcommand{\bb}{\bibitem}
\begin{document}
\title{Multi-sine-Gordon Models}
\author{D. Bazeia$^{a}$, L. Losano$^{a}$, J.M.C. Malbouisson$^{b}$, and  J.R.L. Santos$^{a}$}
\affiliation{\small{$^{a}$Departamento de F\'{\i}sica, Universidade Federal da Para\'{\i}ba, Caixa Postal 5008, 58051-970 Jo\~ao Pessoa, Para\'{\i}ba, Brazil\\
$^b$Instituto de F\'{\i}sica, Universidade Federal da Bahia, 40210-340, Salvador, Bahia, Brazil}}


\begin{abstract}
{This work deals with the presence of defect structures in generalized sine-Gordon models. The models are described by periodic potentials, with substructure having one, two, three or more distinct topological sectors, with multiplicity one, two, three or more, respectively. The investigation takes advantage of the deformation procedure introduced in previous work, which is used to introduce the new models, and to study all the defect structures they may comprise.}
\end{abstract}

\maketitle

\section{Introduction}
In this work we study defect structures in relativistic models described by a single real scalar field, which are of current interest in high energy physics \cite{b1,b2} and in condensed matter \cite{b3,b4}. We focus on static solutions in $(1,1)$ space-time dimensions, with the motivation to present families of models, and the defect structures they may comprise. The models are described by a single real scalar field, engendering discrete symmetry, so we search for defect structures of global nature, known as kinks and lumps.

In high energy physics, kinks and lumps are of interest in several different contexts \cite{b1,b2}, including supergravity, branes with a single extra dimension and cosmology
of branes \cite{sugra,brane,cbrane, etc}. In condensed matter, they may contribute to pattern formation \cite{b3}, and to the description of interfaces among distinct configurations in magnetic systems \cite{b4}. They may also appear in other applications, for instance, in the study of nonlinear excitations in Bose-Einstein condensates \cite{bec}, and in the construction of magnetic apparatus at the nanometric scale \cite{a}.

We use the deformation procedure introduced in Ref.~{\cite{dd}}, and we deform the starting model, which is the $\phi^4$ model engendering spontaneous symmetry breaking, to get to new families of sine-Gordon models. We focus on sine-Gordon models because they are of interest in a diversity of contexts; see, e.g., Refs.~\cite{sg1,sg2}. The procedure is implemented via the deformation function and the parameters it comprises, which leads us to the double and triple sine-Gordon models with analytical solutions, etc. Since the potential of the starting model, the $\phi^4$ model, can be described by a superpotential which is known explicitly, explicit expressions for the superpotential for deformed models can also be derived, and this leads to important informations on the corresponding BPS states \cite{bps}, as well as on the energy and stability of the solutions \cite{bs} of the families of models. 

Most of the families of periodic potentials which will be introduced in this work engender periodic substructure, and can be identified by ${\cal M}=1,2,...$, an integer which describes the multiplicity of the substructure of each potential. In this sense, we will describe a diversity of multi-sine-Gordon models, and the defect structures they may comprise. In particular, the double, triple and other multi-frequency sine-Gordon models are of interest to issues concerning integrability \cite{msg}, and they can also be used to generate braneworld scenarios with a single extra dimension \cite{brane,cbrane,etc}, with substructures that could appear in the form of multi-brane scenarios. The scalar field used to describe the periodic models could also interact with another scalar field, to give rise to defect inside defect \cite{did}, and to allow for the presence of brane with internal structure \cite{bis}. In this sense, the present work provides a variety of multi-sine-Gordon models, and study the presence of topological structure at the classical level, offering results which are of current interest to high energy physics, and to other areas of nonlinear science.

We organize the work as follows: In the next Sect.~\ref{pro} we describe the main steps of the deformation procedure. In Sect.~\ref{mod} we single out the deformation function which will be used throughout the work, and there we study specific sine-Gordon models. In Sect.~\ref{sup} we introduce the superpotentials corresponding to the multi sine-Gordon models studied in the previous Sect.~\ref{mod}, and we calculate the energies corresponding to several specific solutions the models comprise. Finally, in Sect.~\ref{end} we present comments and conclusions.

\section{The procedure}
\label{pro}

We start the investigation with a quick review of the deformation procedure \cite{dd}. We consider two real scalar fields $\chi=\chi(x,t)$ and $\phi=\phi(x,t)$ and the corresponding models, described by the Lagrange densities,
\bes
\ben
{\cal L}=\frac12\partial_\mu\chi\partial^\mu\chi-U(\chi),\label{1a}
\\
{\cal L}_d=\frac12\partial_\mu\phi\partial^\mu\phi-V(\phi).\label{1b}
\een\ees
Here, $U(\chi)$ and $V(\phi)$ are given potentials, which specify each one of the two models. We introduce a function $f=f(\phi),$ named deformation function, from which we link the model (\ref{1a}) with the deformed model (\ref{1b}) by relating the two potentials $U(\chi)$ and $V(\phi)$ in the very specific form
\be
V(\phi)=\frac{U(\chi\to f(\phi))}{(df/d\phi)^2},
\ee
where $U(\chi\to f(\phi))$ means that in the potential $U(\chi)$, we change $\chi$ by $f(\phi)$. In this case, if the starting model has finite energy static solution $\chi(x)$ which obeys the equation of motion
\be
\frac{d^2\chi}{dx^2}=\frac{dU}{d\chi},
\ee
then the deformed model may also have finite energy static solution given by 
\be
\phi(x)=f^{-1}(\chi(x)),
\ee
which obeys
\be
\frac{d^2\phi}{dx^2}=\frac{dV}{d\phi}.
\ee
We note that if the deformation function $f(\phi)$ has critical points, the deformation procedure works smoothly if and only if the potential $U(f(\phi))$ has  zeros of multiplicity at least two at those critical points. The critical points of $f(\phi)$ are the branching points of $f^{-1}(\chi),$ which is consequently a multivalued function.

The deformation procedure has been used by several authors in a diversity of scenarios, and it appears to work correctly in many distinct situations \cite{ddsal,dd+}. 

We go further into the problem using the standard model 
\be\label{vphi4}
U(\chi)=\frac12{(1-\chi^2)^2}.
\ee
This is the $\chi^{4}$ model, with spontaneous symmetry breaking. Here we are using natural units, and we have rescaled the field, and the space and time coordinates to make them dimensionless. This model has as topological defects the BPS states $\chi_{\pm}(x)=\pm\tanh(x-x_0)$, and we choose the center of the solutions at the origin $x_0=0$, for simplicity. The energy density is given by $\varepsilon(x)={\rm sech}^4(x)$, which gives the energy $E=4/3$. We also note that the potential has minima at ${\bar\chi}_\pm=\pm1$ and obeys: $V({\bar\chi}_\pm)=0,$ $V^\prime({\bar\chi}_\pm)=0$, and $V^{\prime\prime}({\bar\chi}_\pm)=4$, where $V^\prime(\chi)=dV/d\chi$, etc. 
Next, if we choose the deformation function in the form $f(\phi)=\cos(\phi)$, we get to the potential
\be\label{equ14}
V(\phi)=\frac12\;{\sin^2(\phi)}.
\ee
In this case the deformation function depends only on $\phi$, and the corresponding deformed model describes the sine-Gordon model, with no extra parameter involved in the procedure. This model has solutions described by 
\be\label{ssg}
\phi_k(x)=\pm\arccos[\pm\tanh(x)]+2k\pi,
\ee
where $k=0,\pm1,\pm2,...$ identifies the particular topological sector of the sine-Gordon model, which has an infinity of sectors. We then see that we go from the $\phi^4$ model, which contains a single topological sector, to the sine-Gordon model, which contains an infinity of topological sectors, by the use of the deformation $f(\phi)=\cos(\phi)$, which is a periodic function, which needs the presence of the integer $k$ to identify the particular periodic sector, which is naturally used to identify the particular topological sector of the deformed model, as we have just shown.

\section{New families of models}
\label{mod}

Here we follow the methodology of Ref.~{\cite{ddsal}}, searching for new functions $f=f(\phi)$, from which we can build families of periodic models. We take the starting model described by the potential \eqref{vphi4} and the kink-like solution $\chi_1(x)=\pm\tanh(x)$, and we choose the deformation function
\be\label{df}
f(\phi)=\cos(a\arccos(g({\phi}))-M\pi).
\ee
This function depends on the choice of $g({\phi})$, on $a,$ which is real constant, and on $M$, which can be integer or semi-integer: for $M$ integer, the deformed potential can be written in the form
\be\label{vs1}
{V}_{\sin}^a(\phi)=\frac{1}{2a^2}\frac{(1-g^2(\phi))}{(g'(\phi))^2}\sin^2\left(a\;\arccos(g(\phi))\right),
\ee
where $g'(\phi)=dg/d\phi$.  If we choose $M$ semi-integer, we get
\be\label{vc1}
{V}_{\cos}^a(\phi)=\frac{1}{2a^2}\frac{(1-g^2(\phi))}{(g'(\phi))^2}\cos^2\left(a\;\arccos(g(\phi))\right).
\ee
As we show below, using integer values for the parameter $a$ lead the number of vacua of the new model to be fixed at will.

The choice \eqref{df} is a generalization of the deformation function used in \cite{ddsal}. There, the deformation function was taken as $g(\phi)=\phi$, and this allowed to generate families of polynomial potentials. In the present work, below we will take $g(\phi)$ as a periodic function of the scalar field, and this will generate new families of models, with periodic potentials, as generalized sine-Gordon potentials. All the potentials are new periodic potentials, and we solve them finding the defect structures they may comprise directly, using the deformation procedure. In this sense, the present work contribute to offer new models, to study the defect structures they may comprise, and to strengthen the power of the deformation procedure introduced in \cite{dd}.
 
In general, the above models identify families of potentials which present static solutions given by
\be\label{sol1}
{\phi}(x)=g^{-1}\left(\cos\left(\frac{\theta(x)+ M\;\pi}{a}\right)\right),
\ee
where $g^{-1}$ represents the inverse of $g(\phi)$. Also, $\theta(x)=\arccos(\tanh(x))$, with $\theta \in [0,\pi]$.
However, to investigate specific models, let us start with \eqref{vs1} and $g(\phi)=\phi$. Here the potentials are, for $a$ odd,
\be\label{vsino}
{V}_{\sin}^a(\phi)=\frac{1}{2a^2}\;\prod_{j=1}^{(a+1)/2} \left(1-\frac{\phi^2}{Z_j^{a\,2}}\right)^2\,,
\ee
and for $a$ even,
\be\label{vsine}
{V}_{\sin}^a(\phi)=\frac12\;\phi^2\;\prod_{j=1}^{a/2} \left(1-\frac{\phi^2}{Z_j^{a\,2}}\right)^2\,,
\ee
where $Z_j^{a}=\cos((j-1)\pi/a)$.

If we start with \eqref{vc1} and $g(\phi)=\phi$, the potentials are, for $a$ odd,
\be\label{vcoso}
{V}_{\cos}^a(\phi)=\frac12\;\phi^2(1-\phi^2)\prod_{j=1}^{(a-1)/2} \left(1-\frac{\phi^2}{Z_j^{a\,2}}\right)^2\,,
\ee
and, for $a$ even,
\be\label{vcose}
{V}_{\cos}^a(\phi)=\frac1{2a^2}\;(1-\phi^2)\;\prod_{j=1}^{a/2} \left(1-\frac{\phi^2} {Z_j^{a\,2}}\right)^2\,,
\ee
where ${Z}_j^{a}=\cos((2j-1)\pi/2a)$.

Here we see that the choice $g(\phi)=\phi$ leads us to polynomial potentials, as already investigated in Ref.~{\cite{ddsal}}. In the present work, however, we want to study periodic potentials, of the sine-Gordon type. 
For this reason, we consider $g(\phi)$ as a periodic function of the scalar field.
If we take $g(\phi)=\cos\phi$, from \eqref{vs1} and \eqref{vc1}, we have the potentials
\be
V^a_{\sin}=\frac1{2a^2}\sin^{2}(a\,\phi)\,
\ee
and
\be
V^a_{\cos}=\frac1{2a^2}\cos^{2}(a\,\phi)\,
\ee
which are sine-Gordon potentials with multiplicity ${\cal M}=1$, with the parameter $a$ now controlling the amplitude and periodicity of the potential. In order to obtain new sine-Gordon potentials with higher multiplicity, in the next subsections we take the function $g(\phi)=\alpha\;\cos\phi$, where $\alpha$ will be chosen appropriately.

\subsection{New {\sl sine} family of models}

Let us now consider the function $g(\phi)=\tilde{Z}_k^{a}\cos\phi$, where $a$ is integer and $\tilde{Z}_k^{a}$ is one of the zeros of the potentials \eqref{vsino}, and \eqref{vsine}, except $\phi=0$, and $\phi=1$. In this case, we have $\tilde Z_k^a= \cos\left(\frac{k}{a}\pi\right)$ with $k=1,2,..\frac{a}{2}-1$, for $a$ even, and $k=1,2,..\frac{a-1}{2}$, for $a$ odd.

The polynomial form of ${V}^{a}_{\sin}$, with its zeros (and multiplicities), for $g(\phi)=\phi$, are known \cite{ddsal}. Thus, performing the deformation $g(\phi)=\tilde Z_k^{a}\cos\phi$  in Eqs. \eqref{vsino} and \eqref{vsine}, we have:
\begin{itemize}
\item for $a>1$ odd:
\be\label{vsinodd}
{V}_{\sin}^{a,k}(\phi)=\frac{1}{2a^2\tilde{Z}_k^{a\,2}}\;\sin^2(\phi) \, \prod_{j=1, j\neq k+1}^{(a+1)/2}
\left(1-\frac{\tilde{Z}_k^{a\,2}}{{Z_j^{a}}^2}\cos^2\phi\right)^2\,,\qquad k=1,2,...,\frac{a-1}{2}\,,
\ee
\item for $a>2$ even:
\be\label{vsineven}
{V}_{\sin}^{a,k}(\phi)=\frac{1}{8}\;\sin^2(2\phi)\, \prod_{j=1,j\neq k+1}^{a/2}
\left(1-\frac{\tilde{Z}_k^{a\,2}}{{Z_j^{a}}^2}\cos^2\phi\right)^2\,,\qquad k=1,2,...,\frac{a}{2}-1\,,
\ee
\end{itemize}
where $Z_{j}^a= \cos\left(\frac{j-1}{a}\pi\right)$.
The $sine$ potential $V^{a,k}_{\sin}(\phi)$ can be written in terms of the Chebyshev
polynomials of second kind:
\bes
\be
V^{a,k}_{\sin}(\phi)=\frac1{2a^2 \tilde{Z}_k^{a\,2}}\;\frac{\left(1-\tilde {Z}_k^{a\,2}\cos^2\phi\right)^2}{\sin^2\phi}\;U^2_{a-1}\left(\tilde{Z}_k^{a}\cos\phi\right)
\ee
\be
U_a(\theta)=\frac{\sin((a+1)\arccos\theta)}{\sin(\arccos\theta)}\label{cheb2}
\ee
\ees
The explicit forms of $V^{a,k}_{\sin}(\phi)$, for $a=3,4,$ and $5$ are given by
\ben\label{vsink}
V^{3,1}_{\sin}(\phi)&=&\frac2{9}\sin^2(\phi)\left(1-\frac14\cos^2(\phi)\right)^2\nonumber\\
&=&\frac{25}{128}\left(\sin(\phi)-\frac1{15}\sin(3\phi) \right)^2\,,
\een
\ben
V^{4,1}_{\sin}(\phi)&=&\frac1{8}\sin^2(2\phi)\left(1-\frac12\cos^2(\phi)\right)^2\nonumber\\
&=&\frac{9}{128}\left(\sin(2\phi)-\frac1{6}\sin(4\phi) \right)^2\,,
\een
\ben
V^{5,1}_{\sin}(\phi)&=&\frac{8\sin^2(\phi)}{25\left(1+\sqrt{5}\right)^2}\left(1-\frac{\left(1+\sqrt{5}\right)^2}{16}\cos^2(\phi)\right)^2\left(1-\frac{\left(1+\sqrt{5}\right)^2}{\left(1-\sqrt{5}\right)^2}\cos^2(\phi)\right)^2\nonumber\\
&=&\frac{\left(483-209\sqrt{5}\right)}{5120}\left(\sin(\phi)-\frac{7\left(5+\sqrt{5}\right)}{2\left(75-37\sqrt{5}\right)}\sin(3\phi)+\frac{\left(7+3\sqrt{5}\right)}{2\left(75-37\sqrt{5}\right)}\sin(5\phi)\right)^2,\nonumber\\
\\
V^{5,2}_{\sin}(\phi)&=&\frac{8\sin^2(\phi)}{25\left(1-\sqrt{5}\right)^2}\left(1-\frac{\left(1-\sqrt{5}\right)^2}{16}\cos^2(\phi)\right)^2\left(1-\frac{\left(1-\sqrt{5}\right)^2}{\left(1+\sqrt{5}\right)^2}\cos^2(\phi)\right)^2\nonumber\\
&=&\frac{\left(483+209\sqrt{5}\right)}{5120}\left(\sin(\phi)-\frac{7\left(5+\sqrt{5}\right)}{2\left(205+93\sqrt{5}\right)}\sin(3\phi)+\frac{\left(3-\sqrt{5}\right)}{2\left(205+93\sqrt{5}\right)}\sin(5\phi)\right)^2,\nonumber\\
\een
which illustrate the new family of models.\\

We note that we have a diversity of sine-Gordon models, which includes variations of the sine-Gordon model, and the double sine-Gordon, triple sine-Gordon, and so on.
We illustrate these models depicting in Fig.~1 the potentials $V^{5,1}_{\sin}$ and $V^{6,1}_{\sin}$, and in Fig.~2, the potentials $V^{7,1}_{\sin}$ and $V^{8,1}_{\sin}$.
\begin{figure}[hb]
\vspace{1cm}
\begin{center}
\includegraphics[{height=4cm,width=6cm,angle=00}]{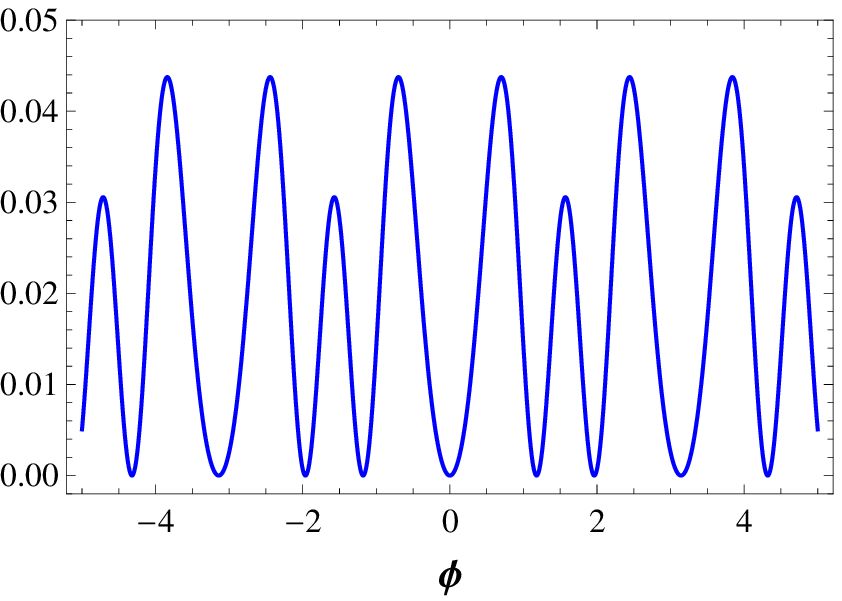}
\hspace{1.6cm}
\includegraphics[{height=4cm,width=6cm,angle=00}]{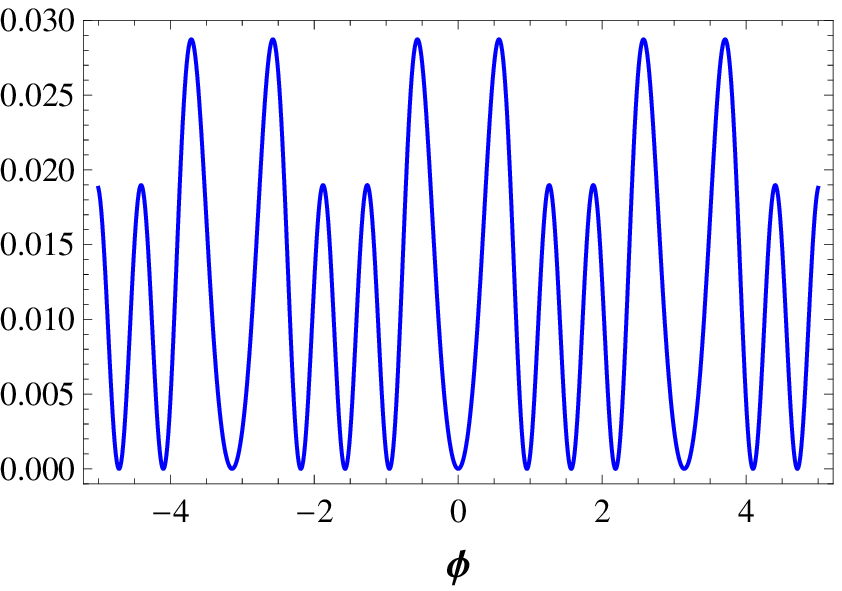}
\end{center}
\caption{Plots of $V^{5,1}_{\sin}(\phi)$, in left panel, and $V^{6,1}_{\sin}(\phi)$, in right panel. These potentials have multiplicity ${\cal M}=2$ and so they belong to the double sine-Gordon family of models.}
\end{figure}
\begin{figure}[ht]
\vspace{1cm}
\begin{center}
\includegraphics[{height=4cm,width=6cm,angle=00}]{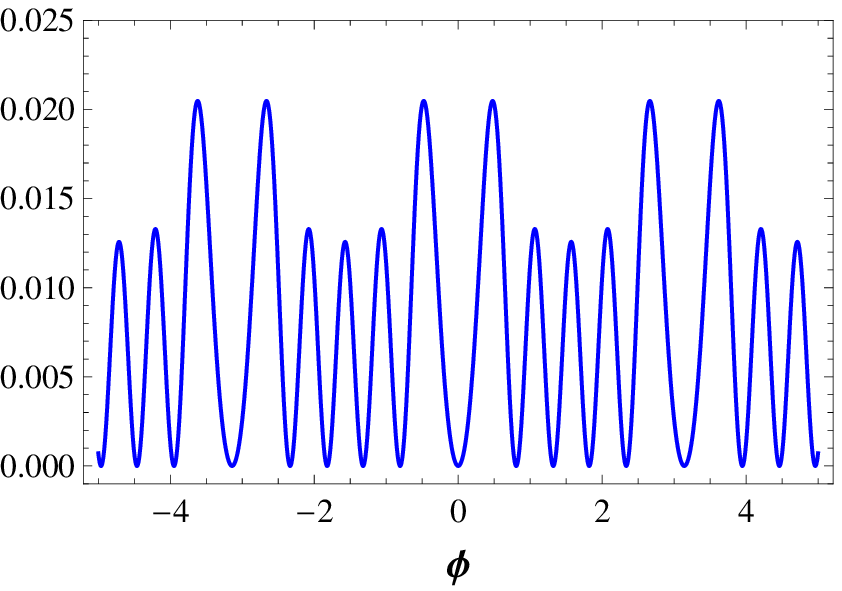}
\hspace{1.6cm}
\includegraphics[{height=4cm,width=6cm,angle=00}]{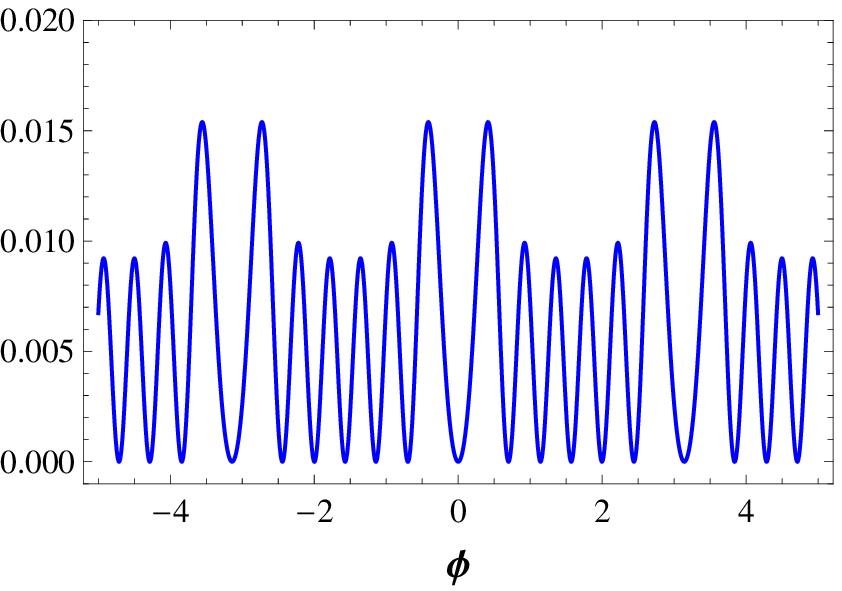}
\end{center}
\caption{Plots of $V^{7,1}_{\sin}(\phi)$, in left panel, and $V^{8,1}_{\sin}(\phi)$, in right panel. These potentials have multiplicity ${\cal M}=3$, and so they belong to the triple sine-Gordon family of models.}
\end{figure}

\subsubsection{Static solutions}

An important advantage of the use of the deformation procedure is that it gives the defect structures explicitly. From \eqref{sol1} and $g(\phi)=\tilde{Z}_k^{a}\cos\phi$, in the case above, we have the static solutions of the $V^{a,k}_{\sin}(\phi)$ given by
\be\label{sols1}
\phi_{m,n}^{a,k}(x)=\pm\arccos\left(\frac1{\tilde{Z}_k^{a}}\cos\left(\frac{\theta(x)}{a}+ \frac{m\;\pi}{a}\right)\right)+n\;{\pi},
\ee
where $\theta(x)=\arccos(\tanh(x))$, $n=0,\pm1,\pm2,..,$ and $m$ is integer in the range: $m=k,...,a-k-1$.

\subsubsection{Maxima and minima}

The maxima and minima of the potentials appear when we take the derivative of these potentials with respect to the field $\phi(x)$ to be zero. 
If the potential is non negative, then the minima are also zeros of it. These informations help us a lot, and in the case of the above potentials, $V_{\sin}^{a,k}$, we can write, for $a$ odd, the minima  
\be
\bar{\phi}_1=n\pi\,,\qquad n=0,\pm 1,\pm 2,...\,,
\ee
and
\be
\bar{\phi}_2=\pm\arccos(\pm Z_{j,k})+2m\pi\,,\;\;\;\; j, k=1,..,\frac{a+1}{2}, j\neq k\,,\qquad m=0,\pm1,\pm2,...
\ee
where $Z_{j,k}={Z_j^a}/{\tilde{Z}_k^a}$. The maxima are 
\be\label{max1a}
\tilde{\phi}_1=\frac{(2n+1)\pi}{2}\,,\qquad n=0,\pm 1,\pm 2,...\,,
\ee
and $\tilde{\phi}_2$, which obeys
\be\label{max1b}
1+\sum_{j=1, j\neq k}^{(a+1)/2}\frac{2Z_{k,j}^{\,2}\sin^2(\tilde{\phi}_2)}{1-Z_{k,j}^{\,2}\cos^2(\tilde{\phi}_2)}=0\,.
\ee
The maxima \eqref{max1a} have the height, $h=1/2a^2\tilde{Z}_k^{a\,2}$, and the height of the others maxima comes from Eq.~\eqref{max1b}, which can be solved numerically. In the table below we show some examples, for several values of $a$ and $k$. We recall that the maxima are periodic, so in the table we show some positive ones, closer to zero, which represent families of maxima.

\begin{center}
\begin{tabular}{|c|c|c|c|c|c|c|c|c|c|}
\hline
 $a$    & $\,3\,$  & $\,4\,$  & $\,5\,$  & $\,5\,$  & $\,6\,$  & $\,6\,$  & $\,7\,$   & $\,7\,$  & $\,7\,$ \\ 
\hline 
 $k$    & $\,1\,$ & $\,1\,$ & $\,1\,$ & $\,2\,$ & $\,1\,$ & $\,2\,$  & $\,1\,$ & $\,2\,$ & $\,3\,$  \\ 
\hline\hline
$max_1$ & $1.571$ & $0.928$ & $0.699$ & $1.571$ & $0.567$ & $0.936$ & $0.479$  & $0.723$ & $1.571$ \\
\hline
$h_1$   & $0.223$ & $0.078$ & $0.044$ & $0.210$ & $0.029$ & $0.074$ & $0.021$  & $0.044$ & $0.206$ \\
\hline
$max_2$ & -       & $2.214$ & $1.571$ & -       & $1.264$ & $2.205$ & $1.065$  & $1.571$ & -       \\
\hline
$h_2$   & -       & $0.078$ & $0.031$ & -       & $0.019$ & $0.074$ & $0.013$  & $0.026$ & -       \\
\hline
$max_3$ & -       & -       & $2.443$ & -       & $1.878$ & -       & $1.571$  & $2.419$ & -       \\
\hline
$h_3$   & -       & -       & $0.044$ & -       & $0.019$ & -       & $0.012$  & $0.044$ & -       \\
\hline
$max_4$ & -       & -       & -        & -      & $2.575$ & -       & $2.077$  & -       & -       \\
\hline
$h_4$   & -       & -       & -        & -      & $0.029$ & -       & $0.013$  & -       & -       \\
\hline
$max_5$ & -       & -       & -        & -      & -       & -       & $2.663$  & -       & -       \\
\hline
$h_5$   & -       & -       & -        & -      & -       & -       & $0.021$  & -       & -       \\
\hline
\end{tabular}
\end{center}

\subsection{New {\sl cosine} family of models}

Here, we consider the function $g(\phi)=\tilde{Z}_k^{a}\cos\phi$, where $a$ is integer and $\tilde{Z}_k^{a}$ is one of the zeros of the potentials \eqref{vcoso}, and \eqref{vcose}, except $\phi=0$ and $\phi=1$. In the case of the $cosine$ family, we have $\tilde Z_{k}^a= \cos\left(\frac{2k-1}{2a}\pi\right)$, with $k=1,..\frac{a}{2}$, for $a$ even, and $k=1,2,..\frac{a-1}{2}$, for $a$ odd.

The polynomial form of ${V}^{a}_{\cos}$, with its zeros (and multiplicities), for $g(\phi)=\phi$, are known \cite{ddsal}. Thus, we  apply the deformation $g(\phi)=\tilde{Z}_k^{a}\cos\phi$  in Eqs. \eqref{vcoso} and \eqref{vcose} to get:

\begin{itemize}
\item for $a>1$ odd:
\be\label{vcosoddk}
{V}_{\cos}^{a,k}(\phi)=\frac{1}{8}\sin^2(2\phi)\,\left(1-\tilde{Z}_k^{a\,2}\cos^2\phi\right) \!\prod_{j=1,j\neq k}^{\frac{a-1}{2}}\!
\left(1-\frac{\tilde{Z}_k^{a\,2}}{{Z_j^{a}}^2}\cos^2\phi\right)^{\!2},\;\;\; k=1,...,\frac{a-1}{2}\,,
\ee
\item for $a$ even:
\be\label{vcosevenk}
{V}_{\cos}^{a,k}(\phi)=\frac{1}{2a^2\tilde{Z}_k^{a\,2}}\sin^2(\phi)\left(1-\tilde{Z}_k^{a\,2}\cos^2\phi\right)\!\!\!\prod_{j=1,j\neq k}^{\frac{a}{2}}\!\!
\left(1-\frac{\tilde{Z}_k^{a\,2}}{{Z_j^{a}}^2}\cos^2\phi\right)^{\!2},\;\;k=1,...,\frac{a}{2}\,,
\ee
\end{itemize}
where $Z_j^a= \cos\left(\frac{2j-1}{2a}\pi\right)$.
The $cosine$ potentials $V^{a,k}_{\cos}(\phi)$ are given by Chebyshev polynomials,
of the first kind:
\bes
\be
V^{a,k}_{\cos}(\phi)=\frac1{2a^2\tilde{Z}_k^{a\,2}}\;\frac{\left(1-\tilde{Z}_k^{a\,2}\cos^2\phi\right)}{\sin^2\phi}\;T^2_{a}\left(\tilde{Z}_k^{a}\cos\phi\right)\,,
\ee
\be
T_a(\theta)=\cos(a\arccos\theta)\label{cheb1}
\ee
\ees
The explicit form of $V^{a,k}_{\cos}(\phi)$, for $a=4$ and $5$ with $k=1$ are given by

\ben
V^{4,1}_{\cos}(\phi)&=&\frac18\frac{\sin^2(\phi)}{(2+\sqrt{2})}\left(1-\frac{(2+\sqrt{2})}{4}\cos^2(\phi)\right)\left(1-\frac{(2+\sqrt{2})}{(2-\sqrt{2})}\cos^2(\phi)\right)^2\nonumber\\
&=&\frac{\left(235-108 \sqrt{2}\,\right)}{4096}\Bigg[1- \frac{\left(744-524 \sqrt{2}\,\right)}{\left(686-451 \sqrt{2}\,\right)}  \cos(2 \phi )+\frac{\left(72-68 \sqrt{2}\,\right)}{\left(686-451 \sqrt{2}\,\right)}  \cos(4 \phi )\nonumber\\
&&-\frac{\left(24+12 \sqrt{2}\,\right)}{\left(686-451 \sqrt{2}\,\right)}  \cos(6 \phi )+\frac{\left(10+7 \sqrt{2}\,\right)}{\left(686-451 \sqrt{2}\,\right)}  \cos(8 \phi )\Bigg]\nonumber\\
\een

\ben
V^{5,1}_{\cos}(\phi)&=&\frac1{8}{\sin^2(2\phi)}\left(1-\frac{(5+\sqrt{5})}{8}\cos^2(\phi)\right)\left(1-\frac{(5+\sqrt{5})}{(5-\sqrt{5})}\cos^2(\phi)\right)^2\nonumber\\
&=&\frac{\left(97-21 \sqrt{5}\,\right)}{4096}\Bigg[1-\frac{\left(46-22 \sqrt{5}\right)}{\left(194-42 \sqrt{5}\right)}  \cos(2 \phi )-\frac{\left(152-56 \sqrt{5}\right)}{\left(194-42 \sqrt{5}\right)}  \cos(4 \phi )\nonumber\\
&&+\frac{\left(21-33 \sqrt{5}\right)}{\left(194-42 \sqrt{5}\right)}  \cos(6 \phi )-\frac{\left(42+14 \sqrt{5}\right)}{\left(194-42 \sqrt{5}\right)}  \cos(8 \phi )+\frac{\left(25+11 \sqrt{5}\right)}{\left(194-42 \sqrt{5}\right)}  \cos(10 \phi )\Bigg]\nonumber\\
\een
which illustrate this new family of models. Here we note that, we have a diversity of sine-Gordon models, which includes variations of the sine-Gordon model, and the double sine-Gordon, triple sine-Gordon, and so on.

We illustrate some potentials in Fig.~3, where we plot $V^{4,1}_{\cos}$ and $V^{5,1}_{\cos}$ and in Fig.~4, where we plot $V^{6,1}_{\cos}$ and $V^{7,1}_{\cos}$.
\begin{figure}[ht]
\vspace{1cm}
\begin{center}
\includegraphics[{height=4cm,width=6cm,angle=00}]{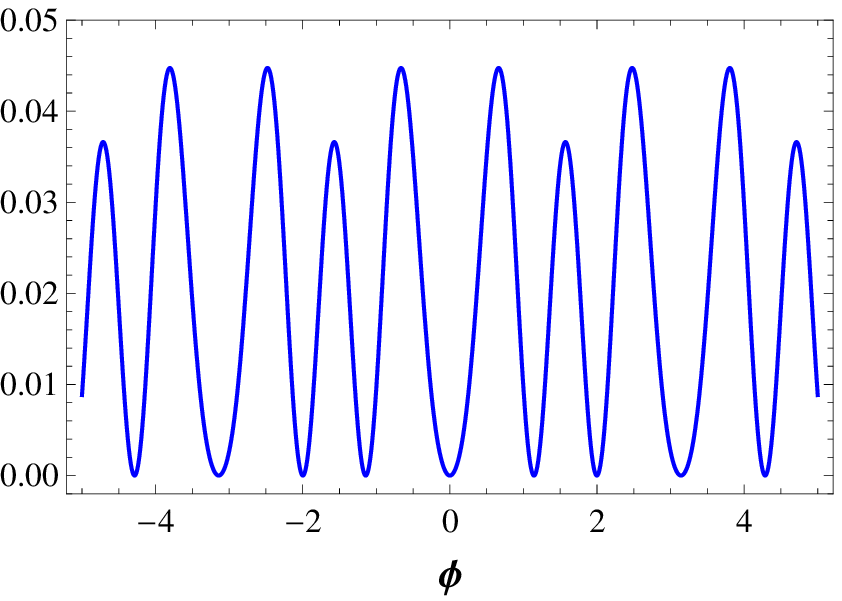}
\hspace{1.6cm}
\includegraphics[{height=4cm,width=6cm,angle=00}]{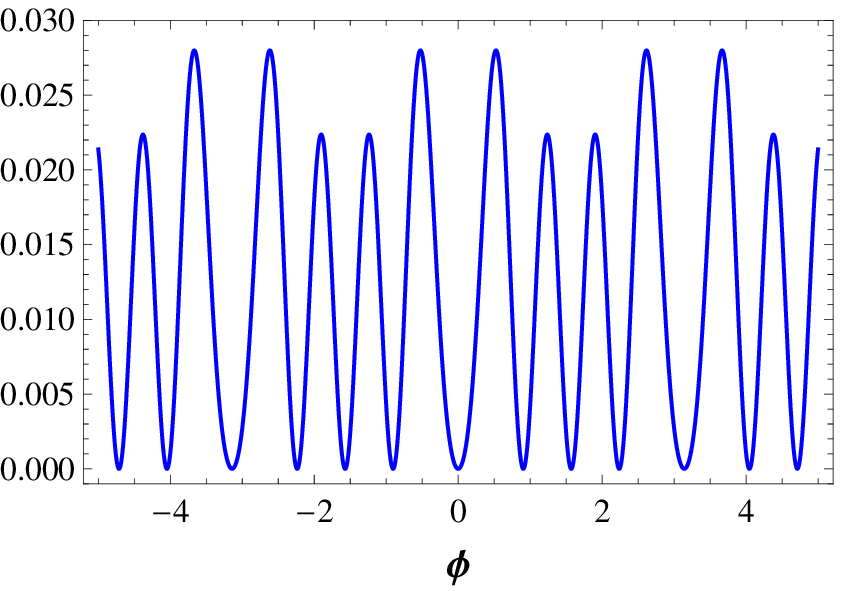}
\end{center}
\caption{Plots of $V^{4,1}_{\cos}(\phi)$, in left panel, and $V^{5,1}_{\cos}(\phi)$, in right panel. These potentials have multiplicity ${\cal M}=2$, so they belong to the double sine-Gordon family of models.}
\end{figure}
\begin{figure}[ht]
\vspace{1cm}
\begin{center}
\includegraphics[{height=4cm,width=6cm,angle=00}]{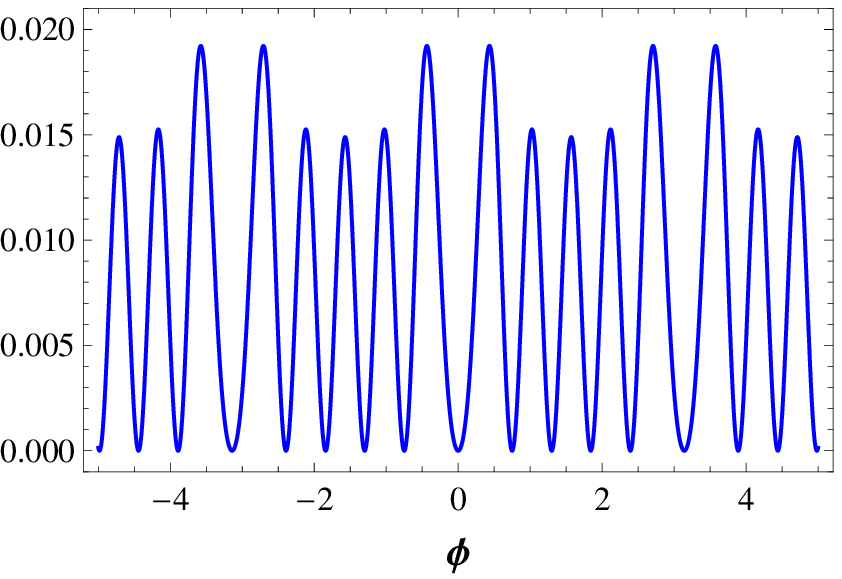}
\hspace{1.6cm}
\includegraphics[{height=4cm,width=6cm,angle=00}]{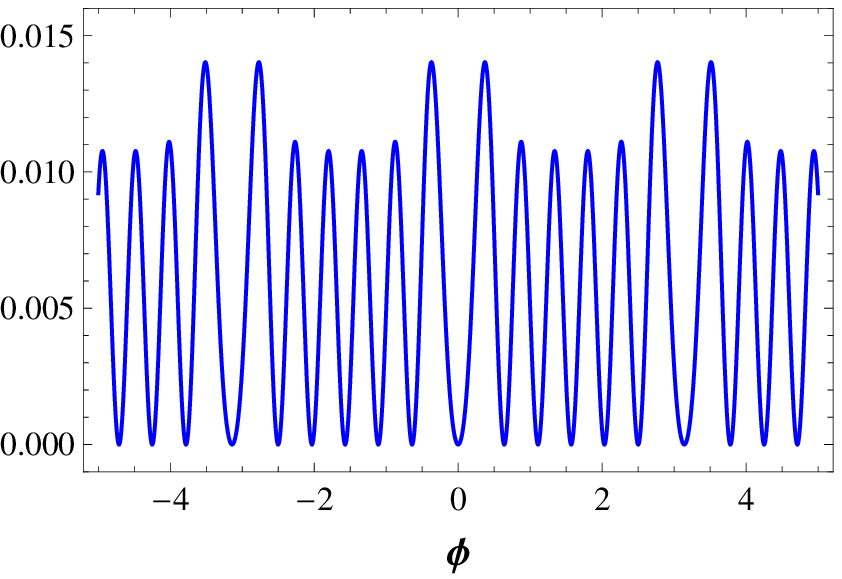}
\end{center}
\caption{Plots of $V^{6,1}_{\cos}(\phi)$, in left panel, and $V^{7,1}_{\cos}(\phi)$, in right panel. These potentials have multiplicity ${\cal M}=3$, so they belong to the triple sine-Gordon family of models.}
\end{figure}

\subsubsection{Static solutions}
We use \eqref{sol1} and $g(\phi)=\tilde{Z}_k^{a}\cos\phi$ to get the static solutions of the $V^{a,k}_{\cos }(\phi)$ family of models. Here we have
\be\label{solc1}
{\phi}_{m,n}^{a,k}(x)=\pm\arccos\left(\frac1{\tilde{Z}_k^{a}}\cos\left(\frac{\theta(x)}{a}+ \frac{(2m-1)\;\pi}{2a}\right)\right)+n\;{\pi},
\ee
where $\theta(x)=\arccos(\tanh(x))$, $n=0,\pm1,\pm2,..,$ and $m$ is integer such that $m=k,...,a-k$.

\subsubsection{Maxima and minima}

Again, like in the case of the $sine$ family, the above potentials are non negative. Therefore,  the minima are calculated taking $V_{\cos}^{a,k}=0$, and from the derivative of the potential, excluding the minima, we obtain the maxima. For $a$ odd, the minima are 
\be
\bar{\phi}_1=\frac{n\pi}{2}\qquad n=0,\pm 1,\pm 2,...\,.
\ee
and
\be
\bar{\phi}_2=\pm\arccos(\pm Z_{j,k})+2m\pi\,,\;\; j, k=1,...,\frac{a-1}{2},j\neq k\,,\qquad m=0,\pm1,\pm2,...\,,
\ee
where $Z_{j,k}=Z_j^a/\tilde{Z}_k^a$.
And, the maxima came from the equation
\ben\label{max4}
&&4\cos(2\tilde\phi)\left(1-{Z_k^a}^{2}\cos^2(\tilde\phi)\right)+{Z_k^a}^{2}\sin^2(2\tilde\phi)+\nonumber\\
&&2\sin^2(2\tilde\phi)\left(1-{Z_k^a}^{2}\cos^2(\tilde\phi)\right)\sum_{j=1,j\neq k}^{(a-1)/2}\frac{Z_{k,j}^{\,2}}{1-Z_{k,j}^{\,2}\cos^2(\tilde\phi)}=0\,,
\een
where $Z_{k,j}=\tilde{Z}_k^a/Z_j^a$, which can be solved numerically, as we did before, for the previous family of models.

\subsection{Kink Multiplicity}

The potential of the sine-Gordon model has an infinity of topological sectors, and supports essencially the same kind of kink. Thus, if ${\cal M}$ denotes the kink multiplicity, as already informed, then for the sine-Gordon model the kink multiplicity is ${\cal M}=1$. In the double sine-Gordon model we have two distinct topological sectors, so the kink multiplicity is ${\cal M}=2$, for the double sine-Gordon model. For the triple sine-Gordon model model we have three distinct topological sectors, so the kink multiplicity is ${\cal M}=3$, and so on. For the periodic models introduced above, we have been able to write the kink multiplicity in terms of the parameter $a$, and in the table below, we summarize the kink multiplicity corresponding to each member of the $sine$ and $cosine$ family of potentials for $a$ integer.

\begin{center}
\begin{tabular}{|l||c|c|}
\hline
Family & $a$ odd  & $a$ even  \\ 
\hline\hline
$V_{\sin}^{a,k}$ & ${\cal M}_k^a=\frac{a+1}{2}-k$ & ${\cal M}_k^a=\frac{a}{2}-k$  \\
\hline
$V_{\cos}^{a,k}$ & ${\cal M}_k^a=\frac{a+1}{2}-k$ & ${\cal M}_k^a=\frac{a+2}{2}-k$  \\
\hline
\end{tabular}
\end{center}

We note that the kink multiplicity depends directly on $a$, so it increases when $a$ also increases.

\section{Superpotentials and energies}
\label{sup}

It is sometimes possible to derive explicit expressions for superpotentials. If $W$ is a function of the scalar field, it is the superpotential if the potential is written in the form
\be\label{sp}
V=\frac12 W^{\prime\,2},
\ee
where the prime stands for the derivative with respect to the field. Now, from the deformation procedure we can write that if $W=W(\chi)$ is the superpotential of the starting model,
then the superpotential ${\cal W}(\phi)$ of the deformed model has to obey
\be
\frac{d{\cal W}}{d\phi}=\frac1{{f^{\prime}(\phi)}}{\frac{dW}{d\chi}\left(\chi\to f(\phi)\right)}.
\ee

In the case of the $\phi^4$ model which we are using to implement the deformation procedure, we have
\be
\frac{dW}{d\chi}=1-\chi^2
\ee
and this can be used to get to the superpotentials of the deformed models.

The expressions of the superpotentials are of great use, since they can be used to get the energies of the corresponding defect structures, since we know that the energy of a topological defect $\chi(x)$ with the potential described by the superpotential $W(\chi)$ is given by
\begin{equation}
E = \left| W(\chi(+\infty)) - W(\chi(-\infty)) \right|.
\end{equation}
where $\chi(\infty)={\rm limit}_{x\to\infty}\chi(x)$, for $\chi(x)$ being the static kinklike solution.

Alternatively, the superpotential can be directly obtained from the integration of Eq.~\eqref{sp},
\begin{equation}
W(\phi) = \pm \int^{\phi} d \xi \, \sqrt{2 V(\xi)}\, ,
\end{equation}
but the integral does not in general lead to an explicit formula for $W$. However, this is not the case here, and the explicit form of the superpotentials corresponding to the $sine$
family, for $a=3,4,5,$ and $6$ are given by
\begin{equation}
{\cal W}_{{\rm sin}}^{3,1}(\phi)=\frac{5}{8} \cos(\phi)-\frac{1}{72}\cos(3 \phi) 
\end{equation}
\begin{equation}
{\cal W}_{{\rm sin}}^{4,1}(\phi)=\frac{3}{16} \cos(2 \phi)-\frac{1}{64}\cos(4 \phi) 
\end{equation}
\begin{equation}
{\cal W}_{{\rm sin}}^{5,1}(\phi) = \frac{(75-37\sqrt{5})}{80(1-\sqrt{5})}
\cos(\phi) - \frac{7(5+\sqrt{5})}{480(1-\sqrt{5})} \cos(3\phi) + \frac{(7+3\sqrt{5})}{800(1-\sqrt{5})}
\cos(5\phi)
\end{equation}
\begin{equation}
{\cal W}_{{\rm sin}}^{5,2}(\phi)\! =\!\frac{1}{160} \!\left(55\!+\!19 \sqrt{5}\right)\!\!\left[ \cos(\phi )-\frac{7}{6}\frac{\left(5+\sqrt{5}\right)}{\left(205\!+\!93 \sqrt{5}\right)} \cos(3 \phi )+\frac{1}{10}\frac{\left(3- \sqrt{5}\right)}{ \left(205\!+\!93 \sqrt{5}\right)} \cos(5 \phi )\right]
\end{equation}
\begin{equation}
{\cal W}_{{\rm sin}}^{6,1}(\phi) = \frac{3}{64}+\frac{11}{256}\cos(2 \phi )+ \frac{3}{64} \cos(4 \phi ) -\frac{3}{256} \cos(6 \phi )
\end{equation}
\begin{equation}
{\cal W}_{{\rm sin}}^{6,2}(\phi) =\frac{1}{64}-\frac{47}{256}\cos(2 \phi )+ \frac{1}{64} \cos(4 \phi ) -\frac{1}{2304} \cos(6 \phi )
\end{equation}

For the $cosine$ family, the explicit form of the superpotentials for
$a=4,$ are given by
\begin{eqnarray}
{\cal W}_{\rm cos}^{4,1}(\phi) & = & \frac{1}{64} \left(\sqrt{2}-1\right) \biggl[16\, \text{arccsc}\left(\frac{2 \sec(\phi )}{\sqrt{2+\sqrt{2}}}\right) \nonumber \\
 & & -\cos(\phi ) \sqrt{5+2 \sqrt{2}-\left(2 \sqrt{2}+3\right) \cos(2 \phi )} \times \left(5 \sqrt{2}-\left(4+3 \sqrt{2}\right) \cos(2 \phi )\right)\biggr]\;\;\;
 \end{eqnarray}
\begin{eqnarray}
{\cal W}_{{\rm cos}}^{4,2}(\phi) & = & \frac{1}{64} \left(\sqrt{2}+1\right) \Bigg[16\,\,\text{arccsc}\left(\frac{2 \sec(\phi )}{\sqrt{2-\sqrt{2}}}\right) \nonumber \\
& &  +\cos(\phi ) \sqrt{5-2 \sqrt{2}+\left(2 \sqrt{2}-3\right) \cos(2 \phi )}\times \left(5 \sqrt{2}+\left(4-3 \sqrt{2}\right) \cos(2 \phi )\right)\Bigg]\;\;\;
\end{eqnarray}

The above expressions allow us to get the energy corresponding to each distinct topological sector. To illustrate this issue, let us calculate some energies. We start with the $sine$ family of models. The potential $V^{3,1}_{\sin}$ has multiplicity ${\cal M}=1$, so it has only one kind of kink, with energy $E=1.222$.
The potential $V^{4,1}_{\sin}$ has multiplicity ${\cal M}=1$, and the kink has energy $E=0.375$. The 
potential $V^{5,1}_{\sin}$ which is plotted in Fig.~1 has multiplicity ${\cal M}=2$, so it has two distinct kinks, which we name large and small kinks, the large one representing the kink in the sector where the maxima of the potential has higher height; here the kink energies are given by $E_{l}=0.211$ and $E_{s}=0.123$. The potential $V^{6,1}_{\sin}$ is plotted in
Fig.~1, and it also has multiplicity ${\cal M}=2$; here the corresponding energies are $E_{l}=0.139$ and $E_{s}=0.076$. Another potential is $V^{7,1}_{\sin}$, which is plotted in Fig.~3. It has multiplicity ${\cal M}=3$, and the corresponding kink energies are $E_{l}=0.099$, $E_{m}=0.053$, and $E_{s}=0.050$, for the large, medium, and small kink, respectively.

Let us now deal with the $cossine$ family of models. The potential $V^{2,1}_{\cos}$ has multiplicity ${\cal M}=1$. The kink energy is $E=1.285$.
The potential $V^{3,1}_{\cos}$ also has multiplicity ${\cal M}=1$, and the kink energy is $E=0.610$. The potential $V^{4,1}_{\cos}$ is shown in Fig.~3,
and it has multiplicity ${\cal M}=2$. Here the two kink energy are $E_{l}=0.206$ and $E_{s}=0.147$, for large and small kink, respectively.
The potential $V^{5,1}_{\cos}$ is plotted in Fig.~3, and it also has multiplicity ${\cal M}=2$. The kink energies are $E_{l}=0.129$ and $E_{s}=0.090$.
Another  potential is $V^{6,1}_{\cos}$, which is depicted in Fig.~4. It has multiplicity ${\cal M}=3$, and the corresponding kink energies are $E_{l}=0.088$, $E_{m}=0.061$, and $E_{s}=0.059$, for large, medium, and small kink, respectively.

\section{Final comments}
\label{end}

In this work we studied a diversity of models described by periodic potentials. The investigation makes use of the deformation procedure, which allows to get to new families of periodic potentials, and to explore all the topological sectors and the corresponding defect structures they may comprise. The periodic potentials present a multiplicity of distinct topological sectors, which we used to define the multi-sine-Gordon models. 

It is worth emphasizing that we have obtained the superpotentials explicitly, which is important to investigate the kink energies. Also, the kink solutions for all the models are known explicitly. In this sense, we believe it is of current interest to investigate how the study of Ref.~{\cite{sg2}} works for the above families of models, and how the results of Ref.~\cite{msg} can be used to the new models introduced in this paper.

The present work opens new lines of investigations. An interesting issue concerns the problem of using the above multi-sine-Gordon models to introduce new braneworld scenarios, in which the brane can engender multistructure. This investigation can be done under the lines of \cite{brane,etc,bis}. These and other related issues are under consideration, and we hope to report on them in the near future.

\acknowledgments

The authors would like to thank CAPES, CNPq, and the Nanobiotec/CAPES program, for partial financial support. 


\end{document}